\documentclass[pre,superscriptaddress,twocolumn,footinbib]{revtex4-1}
\pdfoutput=1
\usepackage[colorlinks=true,urlcolor=black,citecolor=black,linkcolor=black]{hyperref}
\usepackage{amsfonts}
\usepackage{graphicx}
\usepackage{placeins}
\usepackage{amsmath}
\usepackage{xcolor}
\usepackage{color}
\usepackage{bm}
\usepackage[english]{babel}
\usepackage{blindtext}

\definecolor{red}{rgb}{0.75,0,0}
\definecolor{blue}{rgb}{0,0,0.75}
\definecolor{green}{rgb}{0,0.5,0}

\begin{document}

\title{Coupling topological defect phase to extrinsic curvature in nematic shells}

\author{D. J. G. Pearce}
\affiliation{Department of Biochemistry, University of Geneva, 1211 Geneva, Switzerland}
\affiliation{Department of Theoretical Physics, University of Geneva, 1211 Geneva, Switzerland}
\affiliation{NCCR Chemical Biology, University of Geneva, 1211 Geneva, Switzerland}
\affiliation{Dept. of Mathematics, Massachusetts Institute of Technology, Massachusetts, United State of America}

\begin{abstract}

In two dimensional nematics, topological defects are point like singularities with both a charge and a phase. We study topological defects within curved nematic textures on the surface of a cylinder. This allows us to isolate the effect of extrinsic curvature on the structure of the topological defect. By minimizing the energy associated with distortions in the nematic director around the core of a defect we show that the phase of the topological defect is coupled to the orientation of the cylinder. This coupling depends on the relative energetic cost associated with splay, bend and twist distortions of the nematic director. We identify a bistability in the phase of the defects when twist deformations dominate. Finally, we show a similar effect for integer charge topological defects. 

\end{abstract}

\maketitle

\section{Introduction}

Liquid crystals are materials in which the components have a well defined orientation and are able to flow past each other like a fluid~\cite{deGennes:1995,Kleman:2003,Frank:1958}. There are many examples of liquid crystals, from technological displays to biological systems. 

These components of a liquid crystal locally align leading to crystal-like ordering which is described by an orientation field~\cite{deGennes:1995,Kleman:2003,Frank:1958}. Topological defects are topologically protected singularities in this field at which the average orientation is not well defined~\cite{Chaikin:1995,Lavrentovich:2001}. These topological defects are common in nematic phases of liquid crystals, in which the components of the liquid crystal are locally aligned by there is no macroscopic polarization~\cite{Chaikin:1995,Lavrentovich:2001,Kleman:2003,deGennes:1995}. In a two dimensional liquid crystal, topological defects are point like disclinations with typically half integer charge and a phase that is related to their orientation~\cite{Vromans:2016,Tang:2017}. The phase of neighboring defects are coupled by effective elastic torques and forces they exert on each other~\cite{Vromans:2016,Tang:2017,PearceKruse:2021}. 

When confined on a curved surface, there are additional topological and geometrical constraints on the nematic. The topology of the surface constrains the total topological charge of the defects within the nematic~\cite{Hopf:2003}. The intrinsic curvature of the surface can raises or lower the core energy associated with a topological defect depending on its charge~\cite{Giomi:2009}. Finally, the extrinsic curvature exerts a torque directly on the components of the liquid crystal~\cite{Napoli:2012,Napoli:2018}. As a result of this, the extrinsic curvature can exert a force and torque on topological defects within a nematic~\cite{Kralj:2021,Mesarec:2017,Jesenek:2015}.

The interaction between nematics, geometry and topology are increasingly important when considering nematics as a model for biological systems, which are rarely flat and often feature topological defects that correspond to specific geometrical features~\cite{Singer:2016,Saw:2017gn,Kawaguchi:2017,Maroudas:2021,Guillamat:2021,Pearce:2020b}. Furthermore, living systems often feature the continuous injection of mechanical stress, which maintains the system out of equilibrium. Similar systems are referred to as active nematics, in which a defect dense state referred to as ``active turbulence'' is possible~\cite{Ramaswamy:2010,Marchetti:2013,Sanchez:2012,Giomi:2015}. By coupling the system to an externally broken rotational symmetry, it is possible to observe some degree of defect ordering in active nematics~\cite{DeCamp:2015,Putzig:2016,Doostmohammadi:2016,Oza:2016,Pearce:2019,Shankar:2019,Pearce:2020,Thijssena:2020,Pearce:2021,Pearce:2019b,Ellis:2018}.

In this manuscript we investigate how the relative strength of the elastic constants for splay, twist and bend distortions in a nematic lead to an interaction between topological defects and extrinsic curvature. We do this by studying a nematic patch on a cylindrical surface, which allows us to isolate the effects of extrinsic curvature from intrinsic curvature. We begin by reintroducing the general energy of a nematic on a curved surface. We show that when the elastic constants are equal, the equilibrium shape of a topological defect changes to accommodate the extrinsic curvature without changing its phase. However, when the elastic constants are not equal, the phase of the defect becomes coupled to the extrinsic curvature of the surface. This results in topological defects with a fixed phase, which can lead to inter-defect frustration. We identify a region of bistability for $\pm1/2$ charge defects that arises from the effect of twist deformations in the nematic which are identical under $\pi/2$ rotation on a cylinder. Finally we show similar results for $\pm1$ charge topological defects.

\section{Frank free energy on a thin sheet}

A liquid crystal can be described by the unitary director field $\underline{n}$ which denotes the local average orientation of the anisotropic molecules. In the case of a nematic, the molecules are elongated and identical under reversal, thus any physics must be invariant under the transformation $\underline{n} \to -\underline{n}$.

The energy of a nematic liquid crystal depends on variations of the director in space, which can take the form of splay, twist or bend. Thus, the total energy of a nematic texture is given by the Frank free energy \cite{deGennes:1995,Chaikin:1995,Kleman:2003,Frank:1958}
\begin{equation}
\label{eq:Frankenergy}
E_{\rm{F}} = \frac{1}{2}\int_V[K_1(\nabla.\underline{n})^2 + K_2(\underline{n}.\nabla\times \underline{n})^2 + K_3(\underline{n}\times\nabla\times\underline{n})^2 ]\rm{d}V,
\end{equation}
where $K_1$, $K_2$ and $K_3$ are splay, twist and bend elastic constants, respectively. 

When considering a nematic confined to a thin sheet, it is convenient to reduce this equation to a surface energy. Thus we consider a thin nematic layer with uniform thickness $h$ around a regular surface $S$. We assume that $\underline{n}$ lies tangential to the surface at all points and does not vary across the thickness of the sheet. In the limiting case where the sheet thickness is much smaller than the radius of curvature of the sheet, the energy can be written compactly as

\begin{equation}
\label{eq:SFrankenergy}
E_{\rm{F}}^S = \frac{1}{2}\int_S[k_1\kappa_t^2 + k_2\tau_n^2 + k_3(\kappa_n^2 + c_n^2) ]\rm{d}A,
\end{equation}

Where $k_i=hK_i$ are the two dimensional elastic coefficients for splay, twist and bend~\cite{Napoli:2012}. Here we have introduced $\kappa_n$ and $\kappa_t$ as the geodesic curvatures of $\underline{n}$ and its co-normal $\underline{t}$. These correspond to the in-plane bending and splay of the director field, and are not effected by the extrinsic curvature of the surface. We will refer to these terms as the intrinsic energy terms, in curvilinear coordinates, they can be calculated as

\begin{align}
\kappa_n &= \nabla_in^i\\
\kappa_t &= \nabla_i\epsilon^{ij}g_{jk}n^k.
\end{align}
Where $g_{ij}$ is the first fundamental form and $\epsilon^{ij}$ is the Levi-Civita symbol of the surface $S$. These are the terms commonly expected in a two dimensional nematic energy. 

The two additional terms, $\kappa_n$ and $c_n$, refer to the geodesic torsion and normal curvature, respectively. the geodesic torsion gives rise to twist deformations of the nematic, which arise from the fact that on a curved surface, the curl of $\underline{n}$ can have a component tangential to the surface. To minimize this, the geodesic torsion penalizes deviations of $\underline{n}$ from the lines of curvature of the surface. The normal curvature gives rise to bending deformations of the nematic, as the director is confined to the surface which may not be flat. Thus the normal curvature penalizes deviations of $\underline{n}$ from the principal direction of minimal absolute curvature.

The geodesic torsion and normal curvature couple the orientation of the nematic field to the extrinsic curvature of the surface, thus we refer to them as the extrinsic twist and bending energy, respectively. They are given by

\begin{align}
\tau_n &= b_{ij}n^i\epsilon^{jk}g_{kl}n^l\\
c_n &= b_{ij}n^in^j.
\end{align}
Where $b_{ij}$ is the second fundamental form of the surface $S$.

\section{Geometrical setup}

We will consider a small circular patch of nematic with radius $r$ on the surface of a cylinder of radius $R$. It is convenient to use polar coordinates around the center of the patch and we introduce the polar angle $\phi$. We have chosen $\phi=0$ to coincide with the principal curvature of the cylinder, see Fig.~\ref{fig:f1}a. This defines a natural orthogonal basis, $[\underline{e}_\phi,\underline{e}_r]$ in which we shall operate. 

We choose a cylinder here to isolate the effects of extrinsic curvature. Since the surface has no intrinsic curvature, the first fundamental form is identical to that of flat polar coordinates, $g_{\phi\phi}=r^2$, $g_{rr}=1$ and $g_{\phi r}=g_{r\phi}=0$. The second fundamental form captures the extrinsic curvature, and has the form 
\begin{align}
b_{rr} &= \frac{-\cos^2(\phi)}{R}\\
b_{r\phi} &= \frac{r\cos(\phi)\sin(\phi)}{R}\\
b_{\phi\phi} &= \frac{-r^2\sin^2(\phi)}{R}.
\end{align}
Note the complexity here comes from the fact that as $\phi$ changes the single principle curvature appears to rotate. Written in cylindrical coordinates, the second fundamental form would have only one non-zero component reflecting the single, constant principal curvature.

Finally, we can write the unit nematic director as $\underline{n} = n^i\underline{e}_i$ with
\begin{align}
n^r &= \cos(\theta)\\
n^\phi &= \frac{\sin(\theta)}{r}.
\end{align}
Where we have introduced $\theta$ as the angle between the nematic director and the radial basis vector $\underline{e}_r$. Since we are concerned with the forces close to the core of a topological defect, we shall consider the case where $\theta$ is independent of the radial distance, $r$. This allows us to write the terms of the distortion energy as 
\begin{align}
\label{eq:kn}
\kappa_n &= \frac{\cos(\theta)\left[1 + \partial_\phi\theta\right]}{r}\\
\label{eq:kt}
\kappa_t &= \frac{\sin(\theta)\left[1 + \partial_\phi\theta\right]}{r}\\
\label{eq:cn}
c_n &= \frac{-\cos^2(\theta+\phi)}{R}\\
\label{eq:tn}
\tau_n &= \frac{\sin(\theta+\phi)\cos(\theta+\phi)}{2R}.
\end{align}

Writing the terms in this way exposes their nature. The intrinsic bend and splay terms, Eqs.~\ref{eq:kn},\ref{eq:kt}, penalize deviations in the director but are independent of the extrinsic curvature. Whereas the extrinsic twist and bend terms, Eqs.~\ref{eq:cn},\ref{eq:tn}, penalize specific orientations of the director field proportional to the curvature. 

Since we are not considering radial variations in the director field, we calculate this energy for a loop at $r=1$, which we take as our length scale.

\section{Topological defects}

\begin{figure}
	\centering
	\includegraphics[width=\columnwidth]{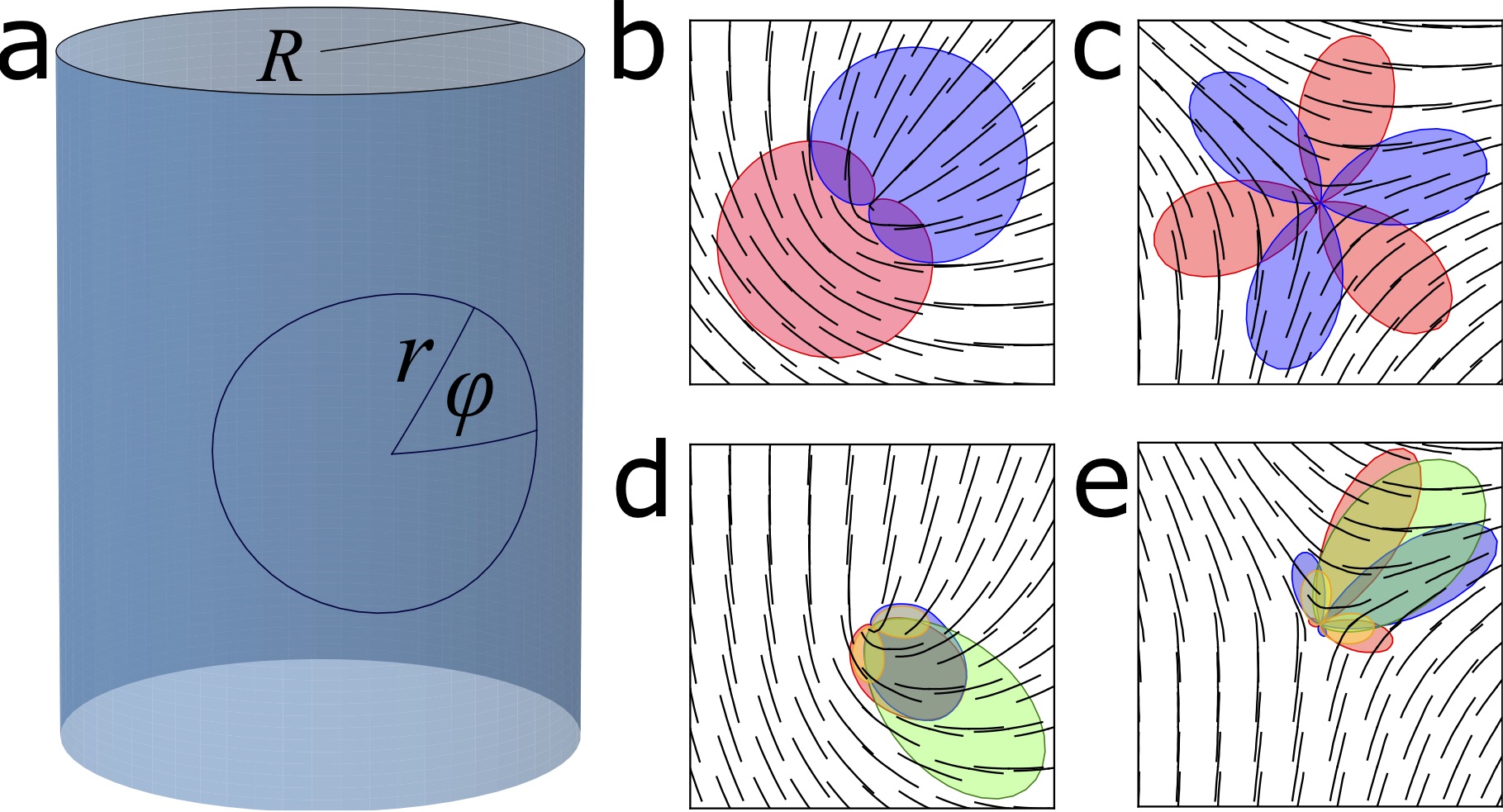}
	\caption{\label{fig:f1} Defects on flat and curved space. (a) Schematic of the coordinate system. The defect is centered at the origin of the polar coordinate system, which exists on the surface of a cylinder with radius R. (b\&c) The equilibrium configuration for a charge (b) $+1/2$ and (c) $-1/2$ topological defect with equal splay, twist and bend constants on a flat surface. The regions of high bend and splay are identified by the red and blue areas, respectively. (d\&e) The equilibrium configuration for a charge (d) $+1/2$ and (e) $-1/2$ topological defect with equal splay, twist and bend constants on the surface of a cylinder. The green and yellow regions show areas with high extrinsic bending and twist, respectively. All defects (b-e) have identical phase, $\psi$.}
\end{figure}

Topological defects in two dimensional nematics are points, such that the director winds by a multiple of $\pi$ on a closed path encircling them~\cite{deGennes:1995,Chaikin:1995,Kleman:2003}. If we place a topological defect at the center of our polar coordinate system, this implies $\oint_0^{2\pi}\frac{\rm{d}\theta}{\rm{d}\phi}\rm{d}\phi = 2\pi (k-1)$, where $k\in\frac{1}{2}\mathbb{Z}$ is referred to as the topological charge of the defect. 

In two dimensional nematics, the lowest energy topological defects have charge $\pm1/2$. For an isolated defect on a flat surface with equal elastic constants and charge $k$ the director field minimizing the Frank free energy is 
\begin{equation}
\label{eq:ndef}
\theta = (k-1)\phi + \psi.
\end{equation}
$\psi$ is referred to as the phase of the defect. Note here that the $k-1$ prefactor is due to the fact that this is written in polar coordinates, which inherently feature a $+1$ topological defect at the origin. Typically, as the charge is increased, the energetic cost of the defect increases linearly, which can be seen by inserting Eq.~\ref{eq:ndef} into Eqs.~\ref{eq:kn},\ref{eq:kt}. For this reason, typically only topological defects with $\pm1/2$ are observed in nematics. Topological defects in nematics have rotational symmetry of order $n=2|k-1|$. Changes in the phase, $\psi$, exchange between bend and splay distortions of the nematic field, and for $k\neq1$ are associated with a global rotation of the defect~\cite{Vromans:2016,Tang:2017}. 

On flat space, energy minimizing topological defects with $k=\pm1/2$ and equal elastic constants are shown in Fig.~\ref{fig:f1}b,c. The shaded regions highlight the magnitude of the bend and splay deformations at each polar angle. The number of alternating peaks of bend and splay distortions around the core of the defect is the same as the order of rotational symmetry. If $k_1=k_3$, the energy is equally distributed around the defect core as both bend and splay. 

We minimize the energy around a defect core using a Monte-Carlo method that preserves the phase of the defect, see appendix for details. This allows us to find the energy minimizing configuration for a given defect phase. We use this method to find the shape of a defect with equal elastic constants on the surface of a cylinder, see Fig.~\ref{fig:f1}d,e. On a curved surface, the defects have lost their symmetry and we see the distortion energy is concentrated into a small angular region. The introduction of extrinsic bending and twisting energy is also visible, as the nematic texture around the defect must accommodate the curvature of the surface. While the curvature of the defects is associated with an increase in the energy, in all cases the degree of intrinsic bend and splay remain balanced. This is because when $k_1=k_3$, the sum of the intrinsic bend and splay energies become independent of the phase, thus independent of the defect orientation relative to the principal curvature of the surface. 

The loss of symmetry around the defect requires a new definition of the phase, which we now calculate as 
\begin{equation}
\label{eq:psi}
\psi = \langle\theta + (1-k)\phi\rangle_\phi.
\end{equation}
Using this definition we see that all defects in Fig.~\ref{fig:f1} have the same phase. It is also compatible with the notion of defect orientation as the angle at which $\theta = 0$, which is in general given by $\phi = \psi/(1-k)$~\cite{Vromans:2016}.

\section{Coupling between defect phase and curvature}

\begin{figure}
	\centering
	\includegraphics[width=\columnwidth]{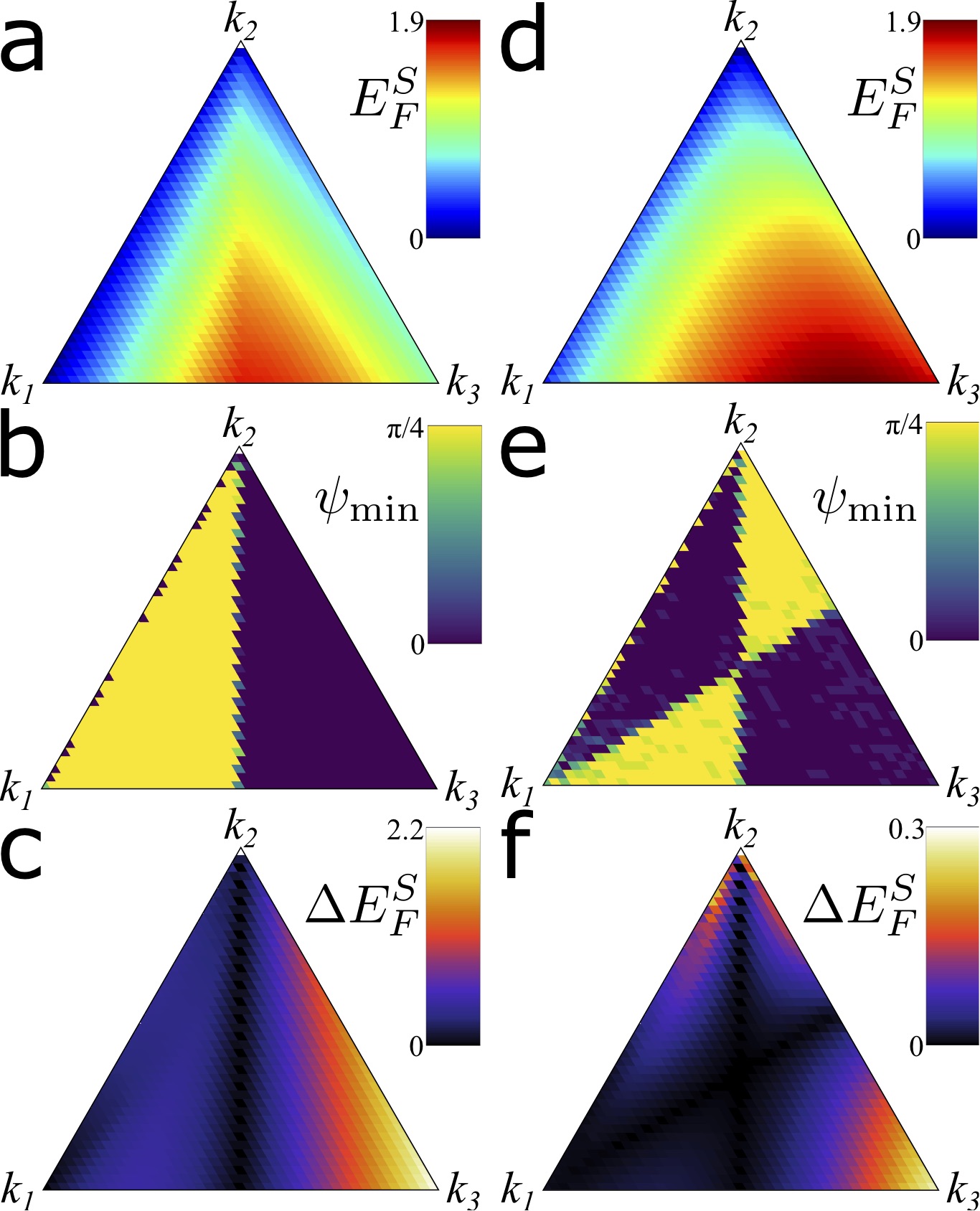}
	\caption{\label{fig:f2} Relative effect of the splay, twist and bend coefficients. (a) Minimum energy cost, $E_F^S$, of a $+1/2$ topological defect on a curved surface. (b) Phase of the energy minimizing $+1/2$ defect, $\psi_{\rm{min}}$. (c) Torque on the $+1/2$ defect core, $\Delta E_F^S$. (d-e) Same as (a-c) but for a $-1/2$ topological defect.}
\end{figure}

We now study the effect of the different elastic constants on defects on a curved surface. We restrict ourselves to the regime $k_1+k_2+k_3 = 1$, since it is only the relative magnitude of the elastic constants that effects the shape of the defect. The energy associated with in-plane distortions depends only on $k_1$ and $k_3$, hence as $k_2$ is increased we see a general reduction in the energy, Fig.~\ref{fig:f2}a,d. This is because as $k_1$ and $k_3$ decrease, the nematic has more freedom to deform in plane to accommodate the curvature. We also see that the energy increases more with $k_3$ than $k_1$, which is expected as $k_3$ corresponds to both intrinsic and extrinsic bending effects. 

Interestingly, we see that if the elastic constants are not equal, there is an energy associated with changes in phase of the topological defect. This leads to a distinct energy minimizing phase which depends on the realtive magnitude of the elastic constants, Fig.\ref{fig:f2}b,e. In the case of a $+1/2$ defect, this effect is anti-symmetric around $k_1=k_3$, Fig.\ref{fig:f2}b. Since the defect phase is associated with the defect orientation in $\pm1/2$ defects, we can also measure an associated defect torque as the energy difference between defects with phase that maximize and minimize the defect core energy, Fig.\ref{fig:f2}c. For $+1/2$ defects the torque disappears at $k_1=k_3$, hence if $k_1\neq k_3$ defects will spontaneously align relative to the cylinder depending on whether the splay or bend coefficient is dominant. 

For $-1/2$ defects the energy minimizing phase is anti-symmetric around $k_1=k_3$ and $k_2=k_3$, leading to four distinct regions with two different phases, Fig.~\ref{fig:f2}e. Again, the transitions between two regions are associated with zero torque on the defects, Fig.~\ref{fig:f2}f. This implies that if $k_1\neq k_3$ and $k_2=k_3$, the phase of $-1/2$ is unaffected by the extrinsic curvature whereas the $+1/2$ defects would experience a torque. This can lead to frustration between defects, as the relative position and phase of defects can also lead to torques on the defect cores~\cite{Vromans:2016,Tang:2017,PearceKruse:2021}. Finally, due to the symmetry of the cylinder, which is identical under $\pi$ rotation, these results are identical under a $\pi/2$ change in phase of the defect, which corresponds to a $\pi$ rotation of the defect~\cite{Vromans:2016}.

\begin{figure}
	\centering
	\includegraphics[width=\columnwidth]{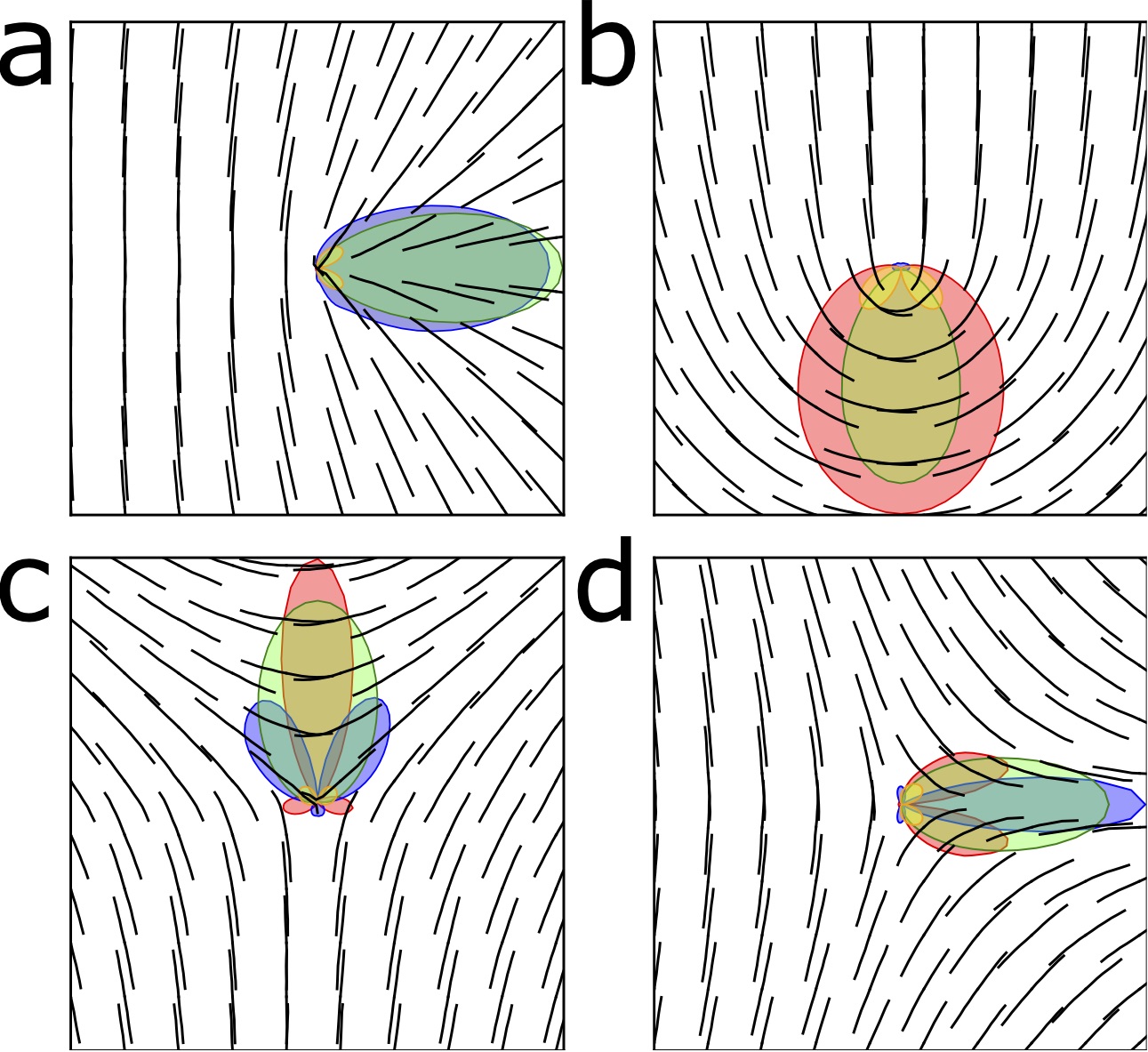}
	\caption{\label{fig:f3} Energy minimizing configurations of half integer topological defects on a cylinder. (a) Bend dominated $+1/2$ defect with phase $\psi=0$.  ($2k_1=2k_2=k_3$) (b) Splay dominated $+1/2$ defect with phase $\psi=\pi/4$. ($k_1=2k_2=2k_3$) (c) Splay and bend dominated $-1/2$ defect with phase $\psi=\pi/4$. ($k_1=10k_2/3=10k_3/7$). (d) Bend dominated $-1/2$ defect with phase $\psi=0$. ($2k_1=2k_2=k_3$).}
\end{figure}

The source of the torques on the defects can be understood by looking at the distribution of energy around the defect core. An interesting property of $+1/2$ defects is that there exist configurations in which the bend (or splay) can be completely removed. These configurations introduce large regions of non-distorted nematic field around the defect core. The extrinsic curvature will cause these regions to align relative to the principal curvatures. This minimizes the extrinsic bend and twist deformations of these sections and results in the torque on the defect core. This is clearly apparent in Fig.~\ref{fig:f3}a,b which show bend and splay minimizing configurations respectively. In both cases, the defect re-orientates such as to minimize the extrinsic bend and twist associated with the parallel region of the nematic. This reorientation is associated with a phase change of the defect hence the defect phase becomes coupled to the extrinsic curvature of the surface. 

The case for $-1/2$ defects is more subtle as there is no bend or splay free configuration of the defect possible. In order to understand the phase-curvature coupling mechanism in $-1/2$ defects we must first consider how the director changes around the core of the defect. As one follows a path around a $\pm1/2$ defect core, the director occupies each orientation once. Since there is one orientation of the director that minimizes $c_n$ and two orientations that minimize $\tau_n$, this implies that at minimum there is one region around the defect core that maximizes $c_n$ and two regions that maximize $\tau_n$. More generally, the number of peaks in the extrinsic bend and twist energy will be $2|k|$ and $4|k|$, respectively. This is visible in Fig.~\ref{fig:f3} where each defect features one green lobe and two yellow lobes corresponding to peaks in the extrinsic bend and twist energy, respectively. By rotating the $-1/2$ defect it is possible to combine the various distortions in different combinations. When the bend distortions have an intermediate energetic cost ($k_1<k_3<k_2$ or $k_2<k_3<k_1$) the defect will re-orientate to combine the bend and extrinsic bend deformations into the same region, which implies combining splay and twist regions, which is only possible when $\psi = \pi/4$, Fig.~\ref{fig:f3}c. However, if the bend distortions have either a high or low cost ($k_3<k_1,k_2$ or $k_3>k_1,k_2$) the defect will re-orientate to separate bend and extrinsic bend, thus combining the extrinsic bend region with splay, and thus twist and bend regions. This is only possible when the defect has phase $\psi=0$, Fig.~\ref{fig:f3}d.

\section{Bistability of defect phase}

\begin{figure}
	\centering
	\includegraphics[width=\columnwidth]{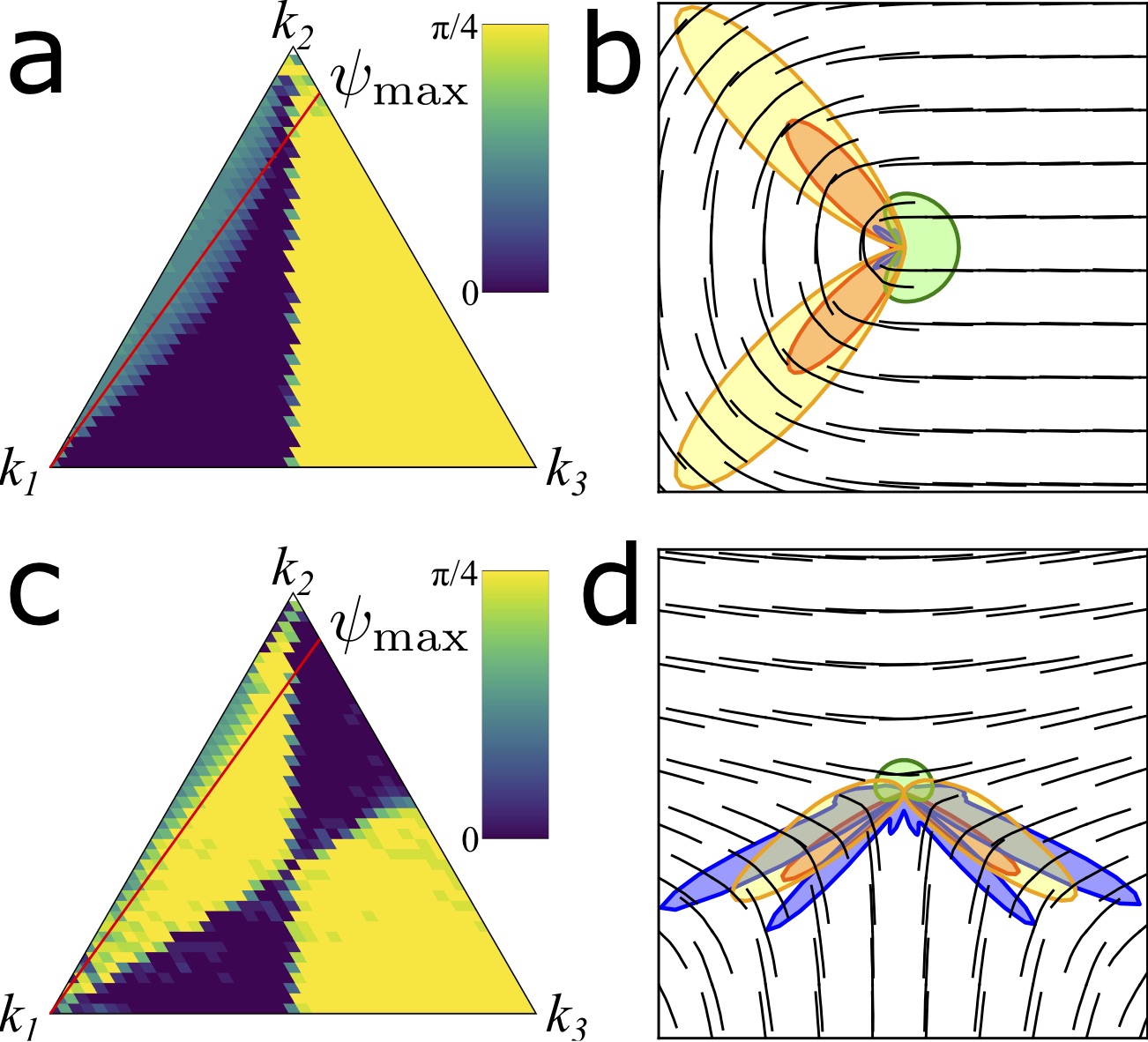}
	\caption{\label{fig:f4} Bistability in defect orientation. (a) Phase of energy maximizing $+1/2$ defect on a curved surface, $\psi_{\rm{max}}$. The red line shows $k_2=8k_3$. (b) Splay and twist dominated stable configuration of a $+1/2$ topological defect with phase $\psi=0$. (c) Same as (a) for $-1/2$ defect. (d) Splay and twist dominated stable configuration of a $-1/2$ topological defect with phase $\psi=\pi/4$.  }
\end{figure}

If we consider the energy of a defect free patch of nematic on the surface of a cylinder, $k=0$, the intrinsic bend and splay energies disappear since $\theta =-\phi + \psi$. Thus the energy density of the patch is given by
\begin{equation}
f = k_3\frac{\cos^4(\psi)}{R^2} + k_2\frac{\sin^2(\psi)\cos^2(\psi)}{4R^2},
\end{equation}
which has fixed points at $\psi = 0$ and $\psi=\pm\pi/2$. Performing a linear stability analysis we see that while $\psi = \pm\pi/2$ is always stable $\psi = 0$ is stable only when $k_2>8k_3$. When $k_3=0$ the system becomes identical under $\pi/2$ rotation and the two stable states are indistinguishable. 

This bi-stability in the defect free nematic texture can lead to bi-stability in the defect phase. To examine this, we check for the phase with the highest energy for $+1/2$ topological defects, Fig.~\ref{fig:f4}a. Largely this looks like the inverse of the energy minimizing phases shown earlier. However, when $k_3$ is very small we see an energy maximizing phase which is neither $\psi=0$ or $\psi=\pi/4$. Since the energy must be symmetric around these values, this implies that there is a bistable state. The additional stable configuration for $+1/2$ defects is splay free and features two peaks in bend energy coincident with regions of high twist, Fig.~\ref{fig:f4}b. This allows for large regions of the defect to be aligned with orientations that minimize the twist distrotion. We see the similar evidence for bi-stability for $-1/2$ defects in a smaller region of the parameter space, Fig.~\ref{fig:f4}c. Similar to the $+1/2$ defect, the $-1/2$ defect features two regions of ordered nematic at orientations that minimize twist deformations seperated by regions of high splay, twist and bend, Fig.~\ref{fig:f4}d.

\section{Integer charge topological defects}

\begin{figure}
	\centering
	\includegraphics[width=\columnwidth]{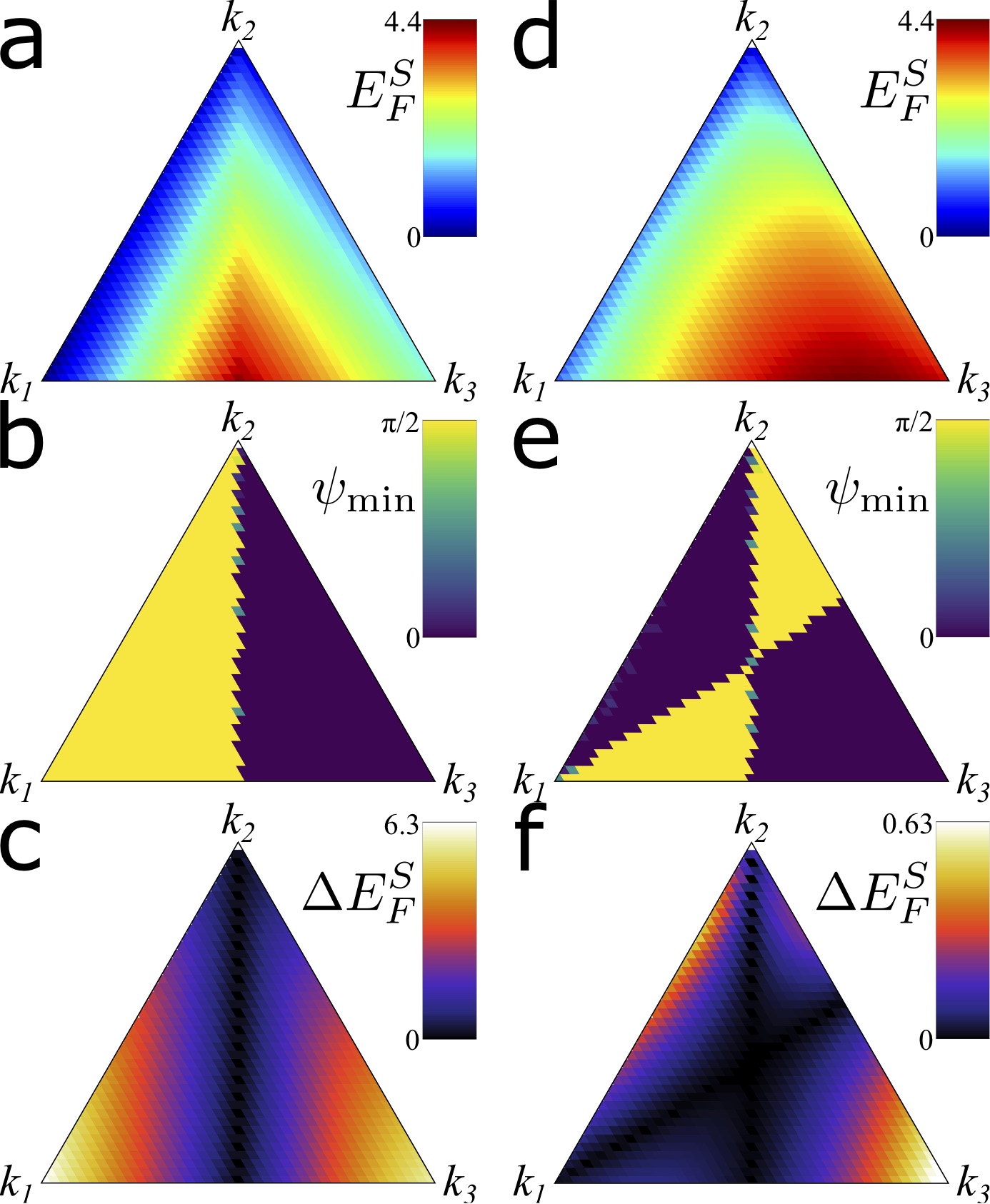}
	\caption{\label{fig:f5}  Relative effect of the splay, twist and bend coefficients. (a) Minimum energy cost of a $+1$ topological defect on a curved surface, $E_F^S$. (b) Phase of the energy minimizing $+1$ defect, $\psi_{\rm{min}}$. (c) Torque on the $+1$ defect core, $\Delta E_F^S$. (d-e) Same as (a-c) but for a $-1$ topological defect. }
\end{figure}

We now look at topological defects with integer charge, $k=\pm1$. This is of particular importance in the study of morphogenesis, where $+1$ topological defects have been shown to induce geometric changes depending on their phase~\cite{Pearce:2020b}. The energy, minimum phase and torque on the defect are very similar to that of a $\pm1/2$ defect, Fig.~\ref{fig:f5}. Again we see a distinction between the positive and negative defects in the energy minimizing phase leading to possible frustration, Fig.~\ref{fig:f5}b,e.

The charge of a topological defect in a nematic can have any half integer number, $k$ with rotational symmetry of order $n=2|k-1|$. When $n$ is odd a phase change of $\Delta\psi=\pi/2$ is equivalent to a reflection of the defect in the $\phi=\psi\pm\pi/2$ line. Since the cylinder is identical under this same reflection, this phase change is not associated with a change in energy. However, if $n$ is even, the topological defect is identical under a reflection in the $\phi=\psi\pm\pi/2$ line. Thus the system is invariant only under a full $\Delta\psi = \pi$ phase change. This effect is visible in Fig.\ref{fig:f5}b,e which now show minimia at $\psi =0$ and $\psi=\pi/2$.

For an integer defect, the same mechanisms identified for $\pm1/2$ defects are present here. For $\pm1$ defect, the director must occupy every orientation at least twice along a path around the defect core. Therefore, there are now two peaks of extrinsic bend and four peaks of twist energy expected around the defect core.

\begin{figure}
	\centering
	\includegraphics[width=\columnwidth]{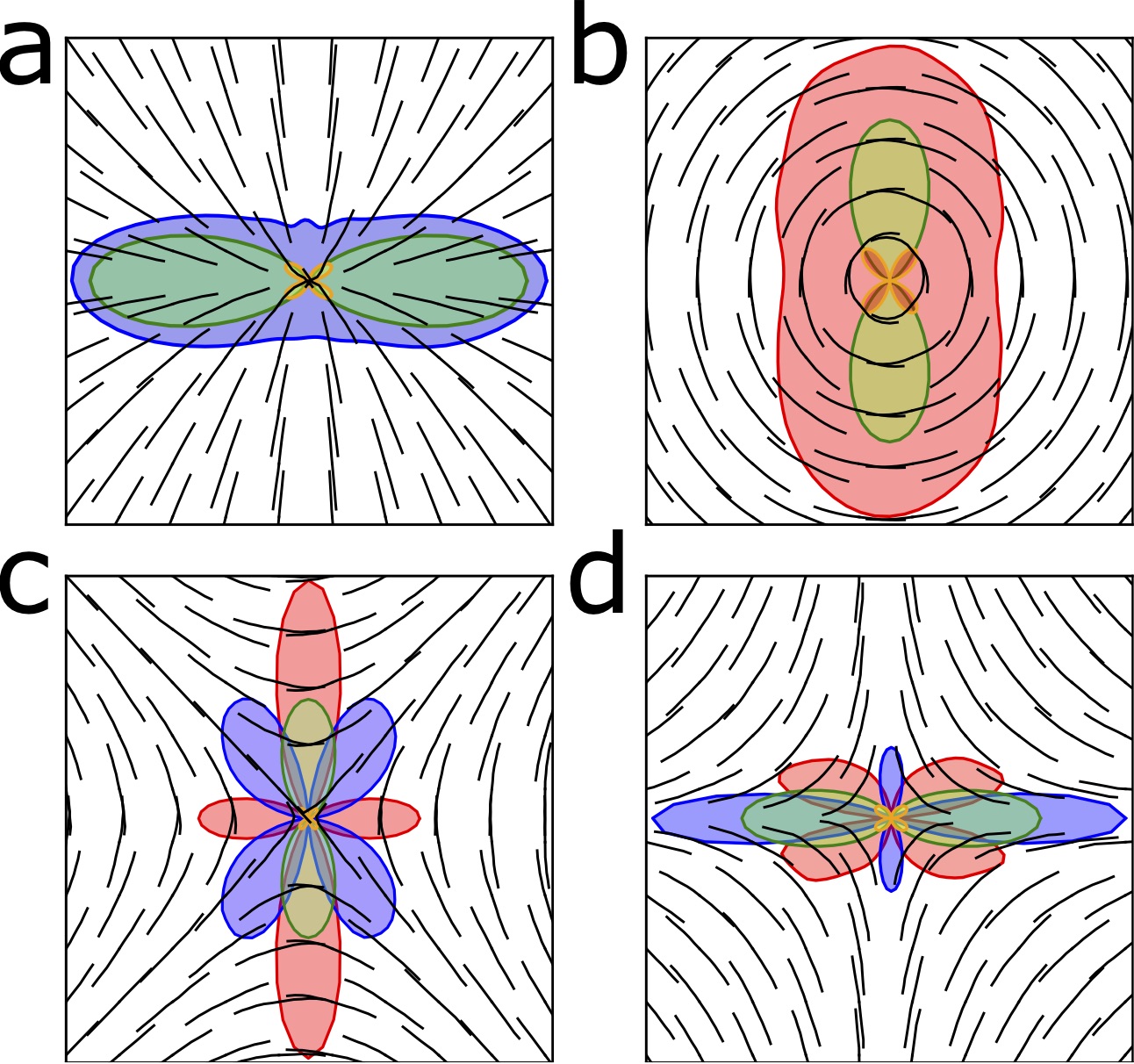}
	\caption{\label{fig:f6} Energy minimizing configurations of integer topological defects on a cylinder. (a) Bend dominated $+1$ defect with phase $\psi=0$.  ($2k_1=2k_2=k_3$) (b) Splay dominated $+1$ defect with phase $\psi=\pi/2$. ($k_1=2k_2=2k_3$) (c) Splay and bend dominated $-1$ defect with phase $\psi=\pi/2$. ($k_1=10k_2/3=10k_3/7$). (d) Bend dominated $-1$ defect with phase $\psi=0$. ($2k_1=2k_2=k_3$). }
\end{figure}

The $+1$ topological defect similar to the $+1/2$ defect in that there exist bend or splay free configurations, however here they are coupled directly to the phase of the defect. For this reason, the $+1$ topological defect has the phase coupled to the elastic constants on a flat surface, where for $k_1<k_3$ we see aster like configurations and we see vortex configurations otherwise~\cite{deGennes:1995,Chaikin:1995,Kleman:2003}. On a curved surface, we see a similar coupling. When $k_1<k_3$ ($k_1>k_3$), the splay (bend) is focused into the two regions where the director aligns with the principal curvature where extrinsic bending is maximized. This allows for large regions to align with the direction of zero extrinsic curvature, Fig.~\ref{fig:f6}a,b.

The $-1$ defect follows a similar mechanism to the $-1/2$ defect. Once again, the defect phase is coupled to the defect orientation. A change in the phase leads to a rotation of the defect, this can be used to align the regions of bend or splay with regions of extrinsic bend or twist. Once again, we see that if $k_2$ is intermediate, the defect will rotate to combine bend and extrinsic bend into the same region of the defect. Otherwise, the defect will rotate to mix bend with twist and splay with extrinsic bend, Fig.~\ref{fig:f6}c,d. For higher charge defects we expect a reduced coupling between the defect phase and curvature. This is because the intrinsic energy of the defect scales with the topological charge $k$, whereas the extrinsic energy of the defect scales with the radius of curvature, $R$ which remains fixed. In addition, the increased number of alternating regions of bend and splay distortion reduce the possibility of distortion-free regions which can align relative to the extrinsic curvature.

\section{Concluding remarks}

By studying topological defects in thin cylindrical nematic sheets, we have been able to demonstrate a coupling between topological defect phase and the extrinsic curvature of the surface which is mediated by the relative elastic constants of the nematic. This effect comes from the interplay between the various types of distortion possible in a nematic. 

When the elastic constants associated with splay, twist and bend distortions are not equal, the distribution of energy within the defect looses symmetry. This loss of symmetry can interact with the broken symmetry of the cylindrical surface leading to a torque on the defect core. This process is independent of many previously discovered mechanisms such as sorting defect charge by Gaussian curvature~\cite{Pearce:2019b,Ellis:2018}, breaking the symmetry of defect generation in active nematics~\cite{Pearce:2020}, or re-orientating defects in the single constant approximation on surface with Gaussian curvature~\cite{Mesarec:2017,Kralj:2021}. 

This work has strong implications in the field of active matter, in particular the development of shape in living systems. Shape changes in living systems are driven by the filamentous cytoskeleton, which has a nematic character and is not generally on a flat plane. Topological defects in the cytoskeleton have been shown to potentially drive morphological changes within growing organisms~\cite{Maroudas:2021,Pearce:2020b}. Thus the placement of the defects is essential for the correct formation and function of the organism. 

\section*{Appendix}
\setcounter{equation}{0}
\def\theequation{A\arabic{equation}}
\subsection{Monte Carlo method for minimizing defect energy.}

We discretize $\phi$ and use a second order finite difference approach to approximate the derivatives in the energy. The defects are initialized with $\theta = (k-1)\phi + \psi$. We perform our energy minimization in a manner that preserves $k$ and $\psi$. 

To minimize the energy by Monte-Carlo method, we define our perturbation:

\begin{equation}
\Delta\theta = \frac{A\sum_{n=-N}^N\exp\left[-\frac{(|\phi-B| +2n\pi)^2}{2C^2}\right]}{\left[ 1 + 2\sum_{n=1}^N\exp[-2n^2\pi^2/C^2]\right]}
\end{equation}

This gives a periodic bump with amplitude $A$, centered on $B$ and width $C$, resulting from the sum of N Gaussians. The parameters are taken from a uniform distribution with limits $A\in[-\pi/4,\pi/4]$, $B\in[0,2\pi]$ and $C\in[0.2,2.2]$. This ensures that the bump has a roughly continuous first derivative at $C+\pi$. We choose $N=1$ to balance computational time against smoothness of the first derivative. 

This deformation does not have the ability to change the topological charge of the defect. In order to maintain a constant phase, we subtract the average integral of the bump.

\begin{equation}
\Delta\theta' = \Delta\theta - \frac{\int_0^{2\pi}\Delta\theta}{2\pi}
\end{equation}

The energy is minimized using a standard simulated annealing process with an exponentially decreasing temperature. $\phi$ is discretized at a resolution of $\rm{d}\phi = 2\pi/100$. For data presented in Fig.~\ref{fig:f2},\ref{fig:f4}a,c\&\ref{fig:f5} we discretize $\psi$ at a resolution of $\rm{d}\psi = \pi/32$. 

\begin{acknowledgments}
I am grateful to Karsten Kruse for insightful discussions. This work was funded by the NCCR for Chemical Biology and the SNSF. 
\end{acknowledgments}

\end{document}